\documentclass[aps,preprint]{revtex4}

\usepackage{amssymb}
\usepackage{amsmath}
\usepackage{epsfig}
\usepackage{epstopdf}
\usepackage{bm}
\usepackage{graphicx,epsfig}
\usepackage{mathrsfs}
\usepackage{dcolumn}
\usepackage{color}
\usepackage{natbib}
\usepackage{CJK}
\def\be{\begin{equation}}
\def\ee{\end{equation}}
\def\bea{\begin{eqnarray}}
\def\eea{\end{eqnarray}}

\allowdisplaybreaks[2]

\begin{document}

\title{Synthetic gauge potential and effective magnetic field in a Raman medium undergoing molecular modulation}
\author{Luqi Yuan$^1$, Da-wei Wang$^2$, and Shanhui Fan$^1$}
\affiliation{$^1$Department of Electrical Engineering, and Ginzton
Laboratory, Stanford University, Stanford, CA 94305, USA \\
$^2$Department of Physics and Astronomy, and Institute for Quantum
Science and Engineering, Texas A\&M University, College Station,
Texas 77843, USA}

\date{\today }

\begin{abstract}
We theoretically demonstrate non-trivial topological effects for a
probe field in a Raman medium undergoing molecular modulation
processes. The medium is driven by two non-collinear pump beams.
We show that the angle between the pumps is related to an
effective gauge potential and an effective magnetic field for the
probe field in the synthetic space consisting of a synthetic
frequency dimension and a spatial dimension. As a result of such
effective magnetic field, the probe field can exhibit
topologically-protected one-way edge state in the synthetic space,
as well as Landau levels which manifests as suppression of both
diffraction and sideband generation. Our work identifies a
previously unexplored route towards creating topological photonics
effects, and highlights an important connection between
topological photonics and nonlinear optics.
\end{abstract}


\maketitle

\begin{figure}[h]
\centering
\includegraphics[width=0.75\linewidth]{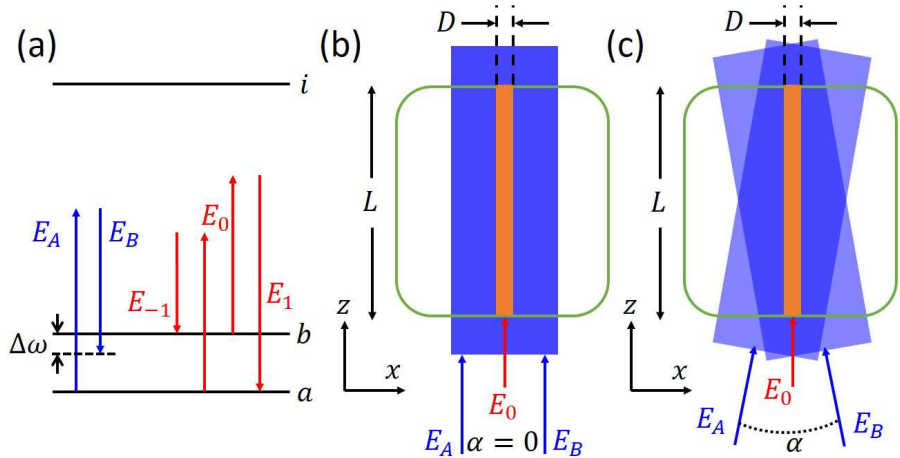}
\caption{(a) Energy levels of a molecule used in the molecular
modulation process. Levels $a$ and $b$ are ground states and $i$
are the excited states. Pump fields $E_A$ and $E_B$ are far from
resonance between the ground states and the excited states, but
are near resonance with the two-photon transition $a$-$b$.
Sidebands $E_n$ are generated via the interaction of molecular
coherence $\rho_{ab}$ as generated from the pump, with the probe
field $E_0$. (b) and (c) Pump fields and the probe field are
injected into the Raman-active medium (inside a cell in the green
square with a length $L$) along near the $z$-direction. Two pump
fields (blue regions) propagate collinearly at an angle $\alpha=0$
in (b) or non-collinearly at $\alpha=0.5^{\circ}$ in (c). The
probe field (red arrow) has a focal area at $z=0$ which is much
smaller than the beam waist of the pump fields. The simulation
region has a width $D$ and is labelled in orange. \label{scheme}}
\end{figure}

The process of molecular modulation (Fig. \ref{scheme}(a)) has
attracted significant interests in the last two decades
\cite{boyd,harris98,sokolov00,sokolov,yzvuz03,huang06,yzvuz07,suzuki08,wang10}.
In this process, molecules are driven by two pump fields, which
generate coherence between a few low-lying vibrational/rotational
levels through a Raman transition. A probe field couples with the
molecular coherence, which results in the generation of Raman
sidebands.  This process is highly efficient and has found
applications in attosecond pulse generation \cite{baker11},
coherent broadband light generation \cite{zhi07,wang14}, and
optical orbital angular momentum transfer
\cite{strohaber12,strohaber15}.

Most previous experiments on molecular modulation assumes a
collinear propagation between the pump and the probe (Fig.
\ref{scheme}(b)). In this Letter, we consider a pump configuration
as shown in Fig. \ref{scheme}(c), where two pump beams are assumed
to be non-collinear, with both their directions near the $z$-axis,
subtending an angle $\alpha \ll 1$ rad. We show that when a probe
beam propagating along the $z$-axis is introduced into this
medium, such a molecular coherence results in a synthetic gauge
field that couples to the probe field. As a result the probe field
exhibits non-trivial topological photonic effect including a
topologically protected one-way edge state along the frequency
axis, as well as Landau levels which manifests as suppression of
both diffraction and sideband generation.

The explorations of synthetic gauge potential
\cite{hafezi11,umucalilar11,fang12prln,fang12np,rechstman12,tzuang14n,li14n,fu15}
and topological effects
\cite{raghu08,haldane08,wang08,khanikaev13,rechstman13,hafezi13,mittal14,lu14,celi14,luo14,wang15}
for light have generated significant recent interests since these
effects open a new dimension in the control of the flow of light.
Most previous works on synthetic gauge field and topological
photonics rely upon complex material geometries. In contrast, in
this work we show that topological photonic effects naturally
arise in a standard nonlinear optics geometry. Our work therefore
points to a potentially fruitful direction at the interface
between nonlinear optics and topological photonics. The predicted
effects also represent a new mechanism for controlling Raman
sidebands in system exhibiting molecular coherence.

We start our analysis by considering a molecular Raman-active
medium. The molecules have a ground state (labelled as ``$a$'' in
Fig. \ref{scheme}(a)), a low-lying excited (labelled as ``$b$''),
and intermediate states (labelled as ``$i$'') at higher energies.
The medium is driven by two pump laser pulses, centered at
frequencies $\omega_A$ and $\omega_B$, respectively. The pumps are
non-resonant with respect to any molecular transitions but are
two-photon near-resonant with the $a$-$b$ transition with a small
detuning $\Delta \omega \equiv (\omega_b - \omega_a)- (\omega_A -
\omega_B)$. The pumps create a coherence $\rho_{ab}$ between
levels $a$ and $b$, which can generate sidebands for the pump
fields themselves \cite{sokolov}. However, we emphasize here that
we are not interested in this sideband generation. Instead we send
a weak probe field $E_0$ into the medium. To distinguish the
different sidebands from the probe and from the pumps, one can
send the probe either as a short pulse that arrives at a time
delay after the short pump pulses have passed through the medium,
or at a very different frequency from the pump fields while the
pumps stay in the medium. In both cases, the key physics for our
purpose here is the interaction of the pump fields with the
coherence generated by the pumps. Therefore, we adopt the analytic
model as was used in \cite{harris98,sokolov,sokolov02}, where both
the pumps and the probe are treated as continuous-wave at a single
frequency, and we provide an estimate on the effects of the pulses
towards the end of the paper.

The propagation equation of a laser beam at a frequency $\omega$
is derived from the Maxwell's equations as
\cite{sokolov,sokolov02}:
\begin{equation}
\frac{\partial}{\partial z} E_{\omega} -i \frac{c}{2\omega}
\frac{\partial^2 }{\partial x^2} E_{\omega}  = i c\mu_0 \hbar
\omega P_\omega. \label{Eqn3:propagation}
\end{equation}
Here, we assume the beam has a propagation direction near the
$z$-axis and use the paraxial wave approximation.  $\mu_0$ is the
vacuum permeability, and $E_\omega$ and $P_\omega$ are the slowly
varying envelopes for the spectral components of the electric
field and the polarization, respectively, at the frequency
$\omega$.

For the pumps $E_{A(B)}$, we use the solution of $P_\omega$ in
Refs. \cite{harris98,sokolov,sokolov02} and write Eq.
(\ref{Eqn3:propagation}) as
\begin{equation}
\frac{\partial}{\partial z} E_{A(B)} -i \frac{c}{2\omega_{A(B)}}
\frac{\partial^2 }{\partial x^2} E_{A(B)} = i  c\mu_0 \hbar
\omega_{A(B)} N \left[ a_{\omega_{A(B)}} \rho_{aa}+
d_{\omega_{A(B)}} \rho_{bb}\right] E_{A(B)} ,
\label{Eqn2:pumpprop}
\end{equation}
where $N$ is the number density of the molecule. In Eq.
(\ref{Eqn2:pumpprop}),
\begin{equation}
a_\omega = \frac{1}{2\hbar^2} \sum_i
\left[\frac{|\mu_{ai}|^2}{(\omega_i-\omega_a)-\omega} +
\frac{|\mu_{ai}|^2}{(\omega_i-\omega_a)+\omega}\right],
\label{Eq1:coeffa}
\end{equation}
\begin{equation}
d_\omega = \frac{1}{2\hbar^2} \sum_i
\left[\frac{|\mu_{bi}|^2}{(\omega_i-\omega_b)-\omega} +
\frac{|\mu_{bi}|^2}{(\omega_i-\omega_b)+\omega}\right],
\label{Eq1:coeffd}
\end{equation}
where $\mu_{a(b)i}$ is the dipole moment between levels $a(b)$ and
intermediate state $i$. In Eq. (\ref{Eqn2:pumpprop}) we only keep
the first-order perturbation to the pumps because we are
interested in the sideband generation from the probe. Because the
pumps are non-resonant with respect to any molecular transition,
we have $a_{\omega_{A(B)}} \cong d_{\omega_{A(B)}} \cong a_0$. By
using $\rho_{aa} + \rho_{bb} =1$, Eqs. (\ref{Eqn2:pumpprop})
becomes
\begin{equation}
i\frac{\partial}{\partial z} E_{A(B)} = - \kappa_{A(B)}
\frac{\partial^2 }{\partial x^2} E_{A(B)}  -\beta_{A(B)} E_{A(B)}
, \label{Eqn2:Epump}
\end{equation}
where $\kappa_{A(B)}=c/2\omega_{A(B)}$, and
\begin{equation}
\beta_{A(B)} = c\mu_0 \hbar \omega_{A(B)} N a_0 .
\label{Eqn10:betaAB}
\end{equation}
Therefore, the pump fields in general can be described as a
Gaussian or Hermite-Gaussian beam with the wave vector
$k_{A(B)}+\beta_{A(B)}$.

The pumps create coherence $\rho_{ab}$ between levels $a$ and $b$,
which oscillates at the frequency $\omega_m = \omega_A - \omega_B$
(see Fig. \ref{scheme}(a)) with an amplitude of
\cite{harris98,sokolov,sokolov02}:
\begin{equation}
\rho_{ab} = \frac{1}{2} \frac{b_{\omega_{A}} E_{A}
E_{B}^*}{\sqrt{|b_{\omega_{A}} E_{A} E_{B}^*|^2 + \Delta
\omega^2}}, \label{Eqn2:coherence}
\end{equation}
where
\begin{equation}
b_\omega = \frac{1}{2\hbar^2} \sum_i
\left[\frac{\mu_{ai}\mu_{bi}^*}{(\omega_i-\omega_a)-\omega} +
\frac{\mu_{ai}\mu_{bi}^*}{(\omega_i-\omega_b)+\omega}\right] .
\label{Eq1:coeffb}
\end{equation}

To study the propagation of the probe, based on the experimental
scenarios as described above, we assume that the coherence does
not decay as the weak probe field propagates through the medium.
The probe has the carrier frequency $\omega_0$. When the probe
interacts with the coherence in the medium, sidebands at
frequencies $\omega_n = \omega_0 + n \omega_m$ are generated,
where $n$ is an integer. From Eq. (\ref{Eqn3:propagation}). the
propagation equation for the electric field in the $n$-th sideband
$E_n$ is \cite{harris98,sokolov,sokolov02}
\begin{equation}
\frac{\partial}{\partial z} E_n -i \frac{c}{2\omega_n}
\frac{\partial^2 }{\partial x^2} E_n = i  c\mu_0 \hbar \omega_n N
\left\{ \left[a_{\omega_n} \rho_{aa}+ d_{\omega_n}
\rho_{bb}\right] E_n + b^*_{\omega_n} \rho_{ab} E_{n-1} +
b_{\omega_{n+1}} \rho_{ab}^* E_{n+1} \right\},
\label{Eq1:maxwell2}
\end{equation}
where $a_\omega$ and $d_\omega$ are defined in Eqs.
(\ref{Eq1:coeffa}) and (\ref{Eq1:coeffd}), and $b_\omega$ is
defined in Eq. (\ref{Eq1:coeffb}). Since all the sidebands are
sufficiently far from any resonance, again we have
\begin{equation}
{\begin{array}{*{20}c}
   a_{\omega_n} \cong d_{\omega_n} \cong a_0; & b_{\omega_n} \cong b_0 \\
\end{array}}  .\label{Eqn:cond}
\end{equation}
Therefore, Eq. (\ref{Eq1:maxwell2}) simplifies to
\begin{equation}
i\frac{\partial}{\partial z}  E_n = \beta_n E_n - \kappa_n
\frac{\partial^2 }{\partial x^2}  E_n - g_n \left( 2 \rho_{ab}
E_{n-1} + 2 \rho_{ab}^*  E_{n+1}\right), \label{Eqn9:Entilde}
\end{equation}
where $\kappa_n=c/2\omega_n$, $\beta_n = c\mu_0 \hbar \omega_n N
a_0$, and $g_n= c\mu_0 \hbar \omega_n N b_0/2$.

From Eq. (\ref{Eqn2:coherence}), with the pump configuration as
described in Fig. \ref{scheme}(c), we have
\begin{equation}
\rho_{ab} (x,z) = \frac{1}{2} \frac{|b_{\omega_{A}} E_{A} (x)
E_{B}^*(x)|}{\sqrt{|b_{\omega_{A}} E_{A}(x) E_{B}^*(x)|^2 + \Delta
\omega^2}} e^{i \theta(x)} e^{-i (\beta_A-\beta_B) z},
\label{Eqn9:coherence2}
\end{equation}
where
\begin{equation}
\theta(x) = q x, \label{Eq1:phase}
\end{equation}
with $q \approx \alpha (k_A + k_B)/2$. On the other hand, from Eq.
(\ref{Eqn10:betaAB}), one can show that $\beta_{n}-\beta_{n-1} =
\beta_m = \beta_A-\beta_B$. Therefore, we perform the
transformation $E_{n} = \tilde E_{n} \exp(i \beta_{n} z)$ and
$\rho_{ab} = \tilde \rho_{ab} \exp(i \beta_{m} z)$ to obtain
\begin{equation}
i\frac{\partial}{\partial z} \tilde E_n = - \kappa
\frac{\partial^2 }{\partial x^2} \tilde E_n - g (x) \left( e^{iq
x} \tilde E_{n-1} + e^{-iq x} \tilde E_{n+1}\right).
\label{Eq:tildeEn}
\end{equation}
In arriving at Eq. (\ref{Eq:tildeEn}), we note that in the limit
of $\omega_0 \gg \omega_m$, $\kappa_n \cong \kappa$ and $2 g_n
|\rho_{ab}| \cong g$ \cite{supple}.

To understand the physics in Eq. (\ref{Eq:tildeEn}), we apply a
gauge transformation $\tilde E_n = \varepsilon_n e^{inq x}$,
define a continuous function $\varepsilon (\omega,x)$ such that
$\varepsilon(\omega_n,x) = \varepsilon_n$, and approximate the
term in the parentheses by a continuous derivative. Eq.
(\ref{Eq:tildeEn}) becomes
\begin{equation}
i\frac{\partial}{\partial z} \varepsilon(\omega,x) \approx \kappa
\left(- i \frac{\partial}{\partial x} +
\frac{\omega-\omega_0}{\omega_m} q \right)^2 \varepsilon(\omega,x)
+ g \omega_m^2 \left(- i \frac{\partial}{\partial \omega}
\right)^2 \varepsilon(\omega,x) - 2g \varepsilon(\omega,x).
\label{Eq:tildevarn}
\end{equation}
Eq. (\ref{Eq:tildevarn}) has the form of a Schr\"{o}dinger
equation in $2+1$ dimensions, except with the usual time axis
replaced by the $z$-axis, and with the remaining two dimensions
describing a \textit{synthetic space} with one spatial dimension
along the $x$ direction and one synthetic frequency dimension
\cite{yuanOL,ozawa16}. In this synthetic space, Eq.
(\ref{Eq:tildevarn}) describes an effective gauge potential
$A_{\omega} = (\omega-\omega_0)q/\omega_m$ along the $x$-axis,
which gives a uniform effective magnetic field orthogonal to the
2D space:
\begin{equation}
B = \frac{\partial A}{\partial \omega} = \frac{q}{\omega_m}
=\alpha\frac{ (k_A + k_B)}{2\omega_m}. \label{Eq:effB}
\end{equation}

\begin{figure}[!h]
\centering
\includegraphics[width=0.8\linewidth]{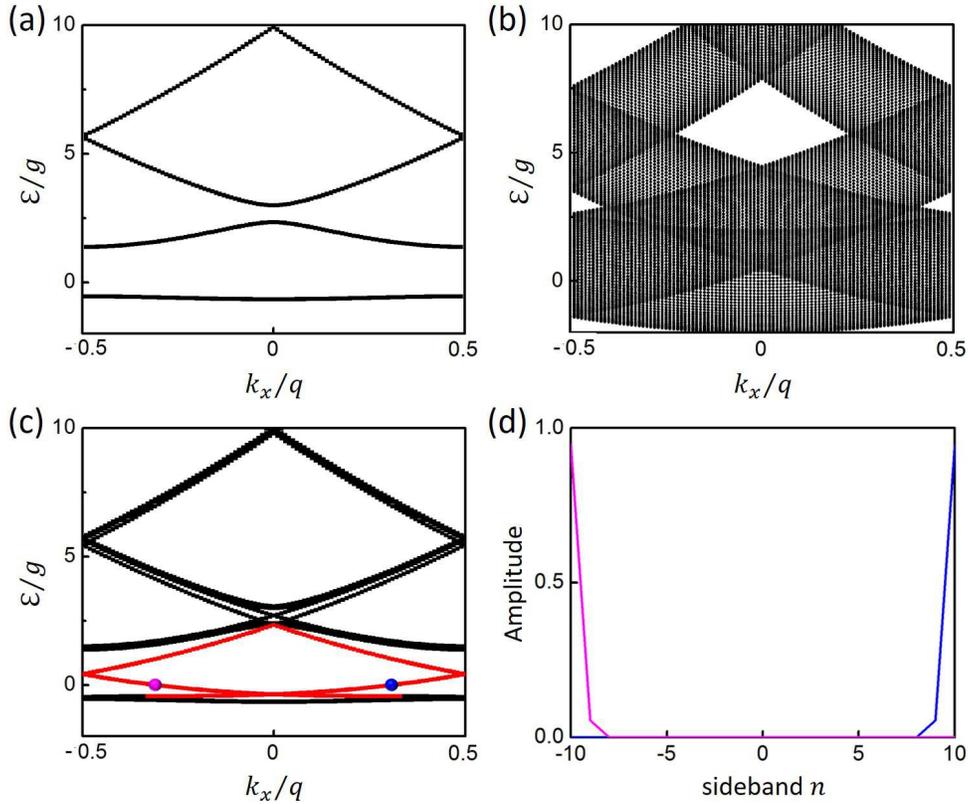}
\caption{(a) The projected bandstructure within the first
Brillouin zone $k_x \in [-q/2,q/2]$ in a infinite system described
by Eq. (\ref{Eq:tildeEn}). We choose $\kappa = g (\pi/2q)^2$. (b)
The projected bandstructure with $q=0$ but with $\kappa$ unchanged
plotted in the same first Brillouin zone. (c) The projected
bandstructure in a 2D strip that is infinite along the $x$-axis
but with a finite number of sidebands (sideband number
$n=-10,\ldots,10$) along the frequency dimension. In the lowest
bandgap, there exists two one-way edge modes (labelled in red).
(d) Field amplitudes of the two edge modes labelled by purple and
blue dots in (c). \label{Fig:band}}
\end{figure}

To examine the topological effect created by such an effective
magnetic field, we calculate the bandstructure of an infinite 2D
system described by Eq. (\ref{Eq:tildeEn}) with a uniform $g(x) =
g$ along the $x$-direction. Eq. (\ref{Eq:tildeEn}) has a spatial
periodicity of $2\pi/q$ along the $x$-axis as well as a symmetry
with respect to the translational operation $n$ to $n+1$ along the
frequency axis. Therefore, it can be described in terms of a
bandstructure $\mathcal{E}(k_x,k_{\omega})$, which relates the
wavevector shift for the probe $\mathcal{E}$ along the
$z$-direction, to the quantum numbers $k_x$ and $k_{\omega}$
corresponding to the translational symmetries as described above.
We take $\kappa = g (\pi/2q)^2$ and plot the projected
bandstructure within the first Brillouin zone $k_x \in [-q/2,q/2]$
in Fig. \ref{Fig:band}(a). Due to the effective magnetic field,
for each $k_x$ the bands are almost completely flat along the
$k_{\omega}$ axis. Therefore, the projected bands appear as lines
in the $\mathcal{E}-k_x$ plane. The bulk bands here correspond to
the Landau level of a particle under a constant magnetic field. In
the continuum limit (such as described by Eq.
(\ref{Eq:tildevarn})) the bands would be completely flat along
both the $k_x$ and the $k_{\omega}$ axis. Here the non-zero slope
along the $k_x$ axis arises from the discrete translational
symmetry. There are gaps between the bands. In contrast, with
$q=0$, which corresponds to the use of two collinear pump beams,
we observe neither the Landau level formation nor the opening of
the band gaps in the projected bandstructure (Fig.
\ref{Fig:band}(b)).

The bands in Fig. \ref{Fig:band}(a) are topologically non-trivial
as characterized by non-zero Chern numbers \cite{tkkn}. Therefore,
in a strip geometry there should be topologically protected
one-way edge states within the gap. As a demonstration, we
consider a strip that is infinite along the $x$-axis but with
finite numbers of $n$ ($n=-10,\ldots,10$) along the frequency
axis. We plot its bandstructure $\mathcal{E}(k_x)$ in Fig.
\ref{Fig:band}(c). In the lowest bandgap, there exists two one-way
edge modes. The field amplitudes corresponding to the two modes at
$\mathcal{E} = 0$ (labelled by purple and blue dots in Fig.
\ref{Fig:band}(c)) are shown in Fig. \ref{Fig:band}(d). One can
see that the fields are located on the two edges and decay
exponentially into the bulk. The bandstructure analysis here
indeed shows the non-trivial topology when the pumps are
non-collinear.

While topological effects have been observed in a wide variety of
photonic systems, the process of molecular modulation provides
unique aspect of probing topological effects. To illustrate these
effects, in what follows we will solve Eq. (\ref{Eq:tildeEn})
numerically for several different pump and probe configurations.
The parameters used in our simulations are based on the recent
experiments in either the gas medium
\cite{sokolov00,sokolov,huang06} or the Raman-active crystal
\cite{zhi07,wang14}. The molecular density is chosen to be $N \sim
10^{18}$-$10^{19}$ cm$^{-3}$. The frequencies of the pump and
probe lasers are at the order of 1 $\mathrm{\mu m}$. Given these
conditions, we have $g \sim 10^{2}$-$10^{3}$ m$^{-1}$, $\kappa
\sim 10^{-7}$ m. At $z=0$, the probe field has a spatial profile
of
\begin{equation}
f(x)=e^{-(x/\Delta x)^2}, \label{Eq:fshape}
\end{equation}
with a focal width $\Delta x=38.8$ $\mathrm{\mu}$m. We will study
the propagation of this probe field along the $z$-axis.

\begin{figure}[h]
\centering
\includegraphics[width=0.7\linewidth]{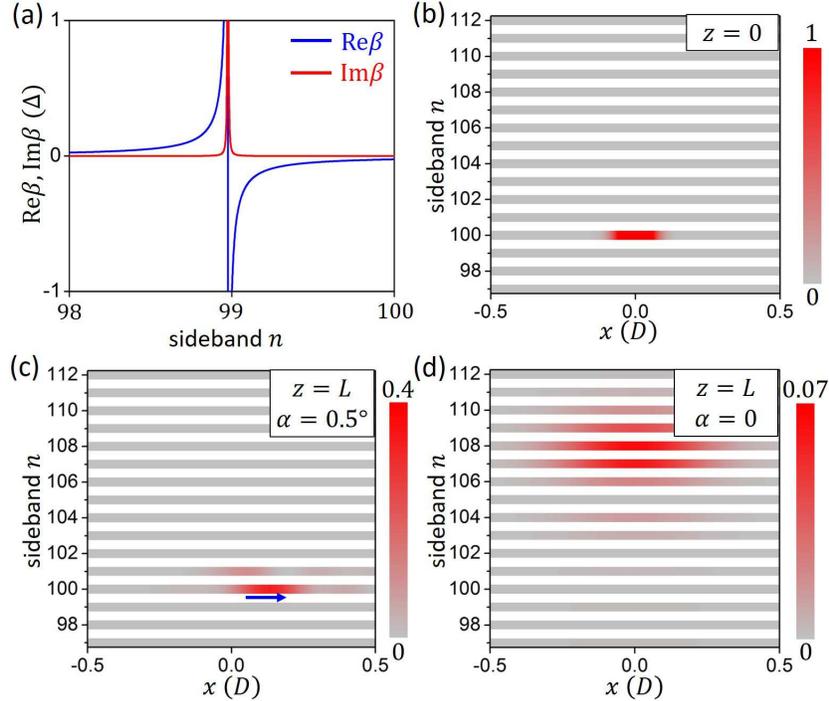}
\caption{(a) The wave vector mismatching for the probe along
$z$-direction, $\beta(\omega_n)$, with an additional two-level
atom added into the medium at a resonant frequency
$\omega_{99}-0.025\omega_m$. $\Delta=10^{3}$ m$^{-1}$.   (b) The
normalized intensity as a function of position and sidebands for
the input probe field. The input field has a frequency of $100.8
\omega_m$. (c) and (d) The output field intensity at $z = L$,
corresponding to the non-collinear pump geometry in Fig.
\ref{scheme}(c) and the collinear pump geometry in Fig.
\ref{scheme}(b). Blue arrow in (c) indicates the propagation
direction of the edge state. \label{Fig:sim2}}
\end{figure}

In Fig. \ref{Fig:sim2} we present the simulation results for the
system with a non-collinear pump geometry with $\alpha=
0.5^{\circ}$, which gives $q = 5 \times 10^{4}$ m$^{-1}$. We
choose $\kappa \sim 10^{-7}$ m, and $g=10^{2}$ m$^{-1}$. These
parameters are the same as used for generating the bandstructure
in Fig. \ref{Fig:band}(c).  The beam waists of the pump fields are
chosen to be $w_0=1$ mm which corresponds to the Rayleigh length
$z_R \sim 3$ m. The length of the medium is $L=5$ cm. We perform
the simulation in a $D\times L$ region as represented by the
orange rectangle in Fig. \ref{scheme}(c), with $D=0.433$ mm,
because the probe field doesn't diffract out of this region in the
entire simulation. Since $D < w_0$ and $L \ll z_R$, we assume that
the amplitudes of the pump fields is uniform in the simulation
region so the coherence is also uniform along $x$-direction in Eq.
(\ref{Eq:tildeEn}). In order to create an edge along the frequency
axis, we add two-level atoms into the system that provides
additional frequency dispersion. We choose two-level atoms to have
a resonant frequency $\omega_{99}-0.025\omega_m$, a density of
$2.5 \times 10^{15}$ cm$^{-3}$, and a dephasing rate $1 / T_2 =
10^{10}/2\pi$ s$^{-1}$. Here $\omega_{99}$ is the frequency of the
$99^{\mathrm{th}}$ sideband, and the $n$-th sideband frequency
satisfies $\omega_n = (n+0.8) \omega_m$.  The wavevector $\beta$
as a function of frequency near the resonant frequency of the
two-level atoms is shown in Fig. \ref{Fig:sim2}(a). Such two-level
atoms strongly influence the wavevectors at the $99^{\mathrm{th}}$
sideband without influencing the wavevector of $100^{\mathrm{th}}$
sideband. With such a choice we expect that the
$100^{\mathrm{th}}$ sideband cannot downconvert, which creates a
boundary along the frequency axis. In the simulation, we input at
$z=0$ a beam at $\omega_{100}$, and we consider 16 sidebands from
$\omega_{97}$ to $\omega_{112}$ (Fig. \ref{Fig:sim2}(b)). After
propagation, the beam shifted towards the $+x$-direction, and
shows very little frequency conversion, in consistency with the
existence of an one-way edge state localized at the lowest
frequency boundary (Fig. \ref{Fig:sim2}(c)). As a comparison, we
study the evolution of the probe with a collinear pump geometry,
i.e. $\alpha=0$, while keeping all the other parameters to be the
same as in Fig. \ref{Fig:sim2}(c) (see Fig. \ref{scheme}(b)). In
the case where the input probe frequency is $100.8 \omega_m $
(Fig. \ref{Fig:sim2}(b)), we observe significant diffraction and
frequency conversion (Fig. \ref{Fig:sim2}(d)). This is consistent
with the theoretical description as presented earlier: in the
collinear pump geometry there is no effective magnetic field and
hence there is no one-way edge state.

\begin{figure}[!h]
\centering
\includegraphics[width=0.7\linewidth]{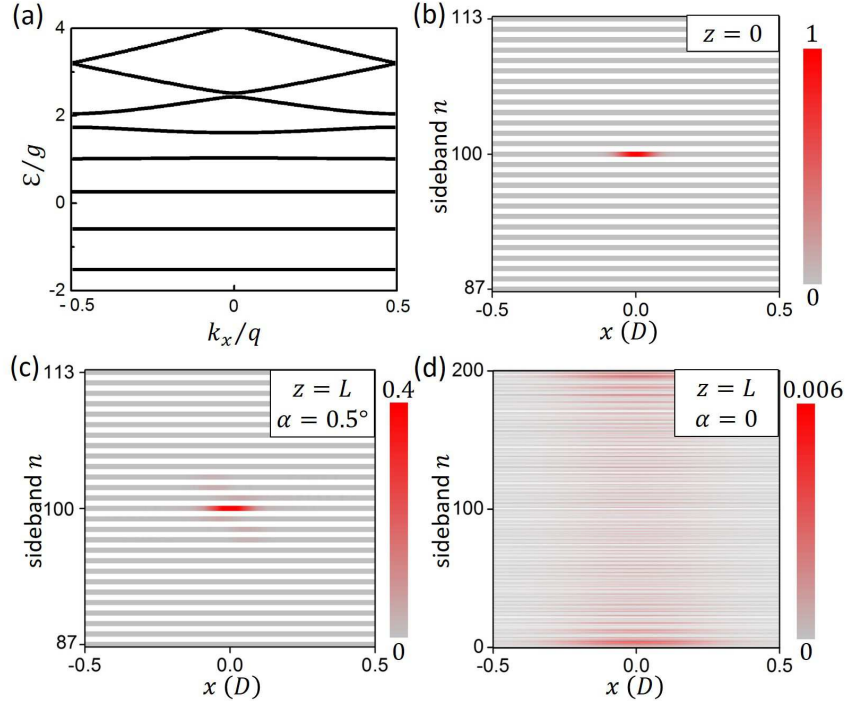}
\caption{(a) The projected bandstructure showing the formation of
Landau levels. $\kappa = 0.1g (\pi/2q)^2$. (b) The normalized
intensity as a function of position and sidebands for the input
probe field. The input field has a frequency of $100.8 \omega_m$.
(c) and (d) The output field intensity at $z = L$, corresponding
to the non-collinear pump geometry in Fig. \ref{scheme}(c) and the
collinear pump geometry in Fig. \ref{scheme}(b). \label{Fig:sim3}}
\end{figure}

For electrons in two dimensions, an important consequence of a
perpendicular magnetic field is the existence of Landau level ---
a bulk band with its energy completely independent of the in-plane
wavevectors. For photons, Landau level has been observed in
\cite{rechstman12}, which relies upon a sophisticated dielectric
geometry. In contrast, here we show that one can directly generate
the Landau level for photons by choosing the right parameters in
systems undergoing molecular modulation (Fig. \ref{Fig:sim3}). As
an illustration, here we choose a larger $g=10^{3}$ m$^{-1}$, and
keep all other system parameters as in Fig. \ref{Fig:sim2}, except
without the additional of the two-level atoms. In general, the
underlying bandstructure of the system, in the absence of the
effective magnetic field, is a tight-binding band along the
frequency axis. The use of a larger $g$ ensures that such
tight-binding band can be approximated by a parabolic band over a
large range of $\mathcal{E}$, which facilitates the creation of
the Landau level in the presence of the effective magnetic field.
Fig. \ref{Fig:sim3}(a) shows the projected bulk band structure
calculation for the system shown in Fig. \ref{scheme}(c), where
the non-collinear pump creates an effective magnetic field. We
indeed observe that the lowest five bands are almost completely
flat in the $k_x$ and $k_\omega$ plane, signifying the creation of
the Landau level. As a demonstration of the effect of the Landau
level, we input the same probe beam as in Fig. \ref{Fig:sim2}, but
with a frequency centered at $100.8 \omega_m$, as shown in Fig.
\ref{Fig:sim3}(b). We see that the probe field does not diffract
in the spatial dimension and also shows no frequency conversion
(Fig. \ref{Fig:sim3}(c)). This is a direct evidence of Landau
levels -- the flattened bands prevent diffraction as well as
frequency conversion. Our system here therefore provides a novel
mechanism to guide light with light. Unlike conventional
waveguides, in which light is guided in a well-defined core
region, here guiding occurs for every spatial and spectral
position inside the ``bulk''. We compare our results with the
evolution of the probe with a collinear pump geometry, i.e.
$\alpha=0$ (Fig. \ref{scheme}(b)). In this case, we choose the
same input probe beam as shown in Fig. \ref{Fig:sim3}(b). Again,
significant diffraction and frequency conversion occurs, as seen
in Fig. \ref{Fig:sim3}(d).

From a theoretical point of view, to observe the effect of the
gauge potential, it is important to keep the diffraction term
(i.e. the term containing $\partial^2/\partial x^2$) in Eq.
(\ref{Eqn3:propagation}) when describing the probe.  Our
theoretical treatment therefore is in contrast with the standard
treatment of molecular modulation process where the diffraction
term is typically ignored.

In the simulations above, we treat the probe field in Eq.
(\ref{Eq:tildeEn}) as a monochromatic field. Our results are also
valid if the pumps and the probe are pulses, as long as the
temporal duration of the pulses are long such that the
slowly-varying-envelope approximation is valid. Using a long pump
pulse, the coherence can be prepared via the ``adiabatic
following'' scheme \cite{eberlybook}, in which a large coherence
can be achieved by setting a negative $\Delta \omega$ and by
adiabatically increasing the strength of the pump fields until
$|\rho_{ab}| = 1/2$ is reached. We give an example of the typical
time scale of pulses in a possible experiment. For the pump one
can use a $\sim 500$ fs pulse, which has a the spectral width of
$\delta\omega_{\mathrm{FWHM}}/\omega_{\mathrm{pump}} \sim
10^{-3}$. For such a pump one can assume that $q$ is a constant in
Eq. (\ref{Eq1:phase}). A typical molecular modulation process
requires a dephasing time of the coherence long enough so it works
in the transient regime before the coherence has time to decay
\cite{sokolovbook}. Therefore, to study the interaction of a probe
pulse with the coherence, one can send in the probe pulse at a
time delay. For example, one can use a $\sim 500$ fs probe pulse
at a time delay $\sim 1$ ps with respect to the pump.
Alternatively, one can also use the long pump pulses $\sim 1$ ns
and send the probe into the medium while the pumps are on. In this
case, one can choose the frequencies of the pump and the probe to
be sufficiently different and observe the predicted topological
effects in the sidebands associated with the probe.

In our simulation, we consider the diffraction effects along only
one spatial dimension (the $x$-dimension). One can choose the
spatial profile of the beam such that the diffraction effect along
the $y$-direction to be sufficiently weak. On the other hand, by
including diffraction effects along both the $x$ and
$y$-dimensions, one may also explore three-dimensional gauge field
physics \cite{lu16,slobozhanyuk16}.

In summary, we study the process of the coherent Raman sideband
generation by the molecular modulation with a non-collinear pump
geometry. We show that such a geometry provides a synthetic gauge
potential for a probe beam sent into the same system. The gauge
potential can introduce non-trivial topological effects for the
probe beam.  Our work identifies a previously unexplored route
towards creating topological photonics effects, and highlights an
important connection between topological photonics and nonlinear
optics.

\begin{acknowledgments}
The authors acknowledge stimulating discussions with Professor
Steve Harris. This work is supported by U.S. Air Force Office of
Scientific Research Grant No. FA9550-12-1-0488.
\end{acknowledgments}


\begin{thebibliography}{99}
\bibitem{supple} From Eq.
(\ref{Eq1:phase}), Eq. (\ref{Eq:tildeEn}) is valid when $q \ll
k_n$. This condition is satisfied in this letter.
\end{thebibliography}
\end{document}